\documentclass{article}

\usepackage{PRIMEarxiv}

\usepackage[utf8]{inputenc} 
\usepackage[T1]{fontenc}    
\usepackage{hyperref}       
\usepackage{url}            
\usepackage{booktabs}       
\usepackage{amsfonts}       
\usepackage{nicefrac}       
\usepackage{microtype}      
\usepackage{lipsum}
\usepackage{fancyhdr}       
\usepackage{graphicx}       
\graphicspath{{media/}}     

\usepackage{color}
\usepackage{xcolor}
\usepackage{listings}
\usepackage{lineno}

\definecolor{codegreen}{rgb}{0,0.6,0}
\definecolor{codegray}{rgb}{0.5,0.5,0.5}
\definecolor{codepurple}{rgb}{0.58,0,0.82}
\definecolor{backcolour}{rgb}{0.95,0.95,0.92}

\lstdefinestyle{mystyle}{
    backgroundcolor=\color{backcolour},
    commentstyle=\color{codegreen},
    keywordstyle=\color{blue},
    numberstyle=\tiny\color{codegray},
    stringstyle=\color{codepurple},
    basicstyle=\ttfamily\scriptsize,
    numbers=left,
    numbersep=5pt,
    breaklines=true,
    breakatwhitespace=true,
    showstringspaces=false,
    captionpos=b,
    frame=single
}

\title{GuardRails: Automated Suggestions for Clarifying Ambiguous Purpose Statements
}

\author{
  Mrigank Pawagi \\
  Indian Institute of Science \\
  Bengaluru\\
  \texttt{mrigankp@iisc.ac.in} \\
   \And
  Viraj Kumar \\
  Indian Institute of Science \\
  Bengaluru\\
  \texttt{viraj@iisc.ac.in} \\
}

\begin{document}
\maketitle

\begin{abstract}
Before implementing a function, programmers are encouraged to write a \emph{purpose statement} i.e., a short, natural-language explanation of what the function computes. A purpose statement may be \emph{ambiguous} i.e., it may fail to specify the intended behaviour when two or more inequivalent computations are plausible on certain inputs. Our paper makes four contributions. First, we propose a novel heuristic that suggests such inputs using Large Language Models (LLMs). Using these suggestions, the programmer may choose to clarify the purpose statement (e.g., by providing a \emph{functional example} that specifies the intended behaviour on such an input). Second, to assess the quality of inputs suggested by our heuristic, and to facilitate future research, we create an open dataset of purpose statements with known ambiguities. Third, we compare our heuristic against GitHub Copilot's Chat feature, which can suggest similar inputs when prompted to generate unit tests. Fourth, we provide an open-source implementation of our heuristic as an extension to Visual Studio Code for the Python programming language, where purpose statements and functional examples are specified as docstrings and doctests respectively. We believe that this tool will be particularly helpful to novice programmers and instructors.
\end{abstract}


\section{Introduction}
Large Language Models (LLMs) can generate code from natural language prompts~\cite{chen2021}. Although this code is not always accurate, an early study showed that LLM-generated code outperforms novice programmers on simple code-writing tasks~\cite{finnie2022robots}. More recent work shows continued improvement, including on more complex programming tasks~\cite{li2022competition, openai2023gpt4}. As a result, there have been calls to urgently review ``our educational practices in the light of these new technologies''~\cite{becker_sigcse23}. One such review of code-writing tasks has been put forward by Raman and Kumar~\cite{raman2022compute}, based on the 6-step recipe proposed by Felleisen et al.~\cite{felleisen2018design} to help novice programmers design functions. We focus on two of these steps:
\begin{description}
\item[Step 2 (Signature, Purpose Statement, Header)] State what kind of data the desired function consumes and produces. Formulate a concise answer to the question \emph{what} the function computes. Define a stub that lives up to the signature.
\item[Step 3 (Functional Examples)] Work through examples that illustrate the function’s purpose.
\end{description}

\subsection{Motivating Example}
\label{sec:motivation}
\begin{figure}
  \includegraphics[width=\linewidth]{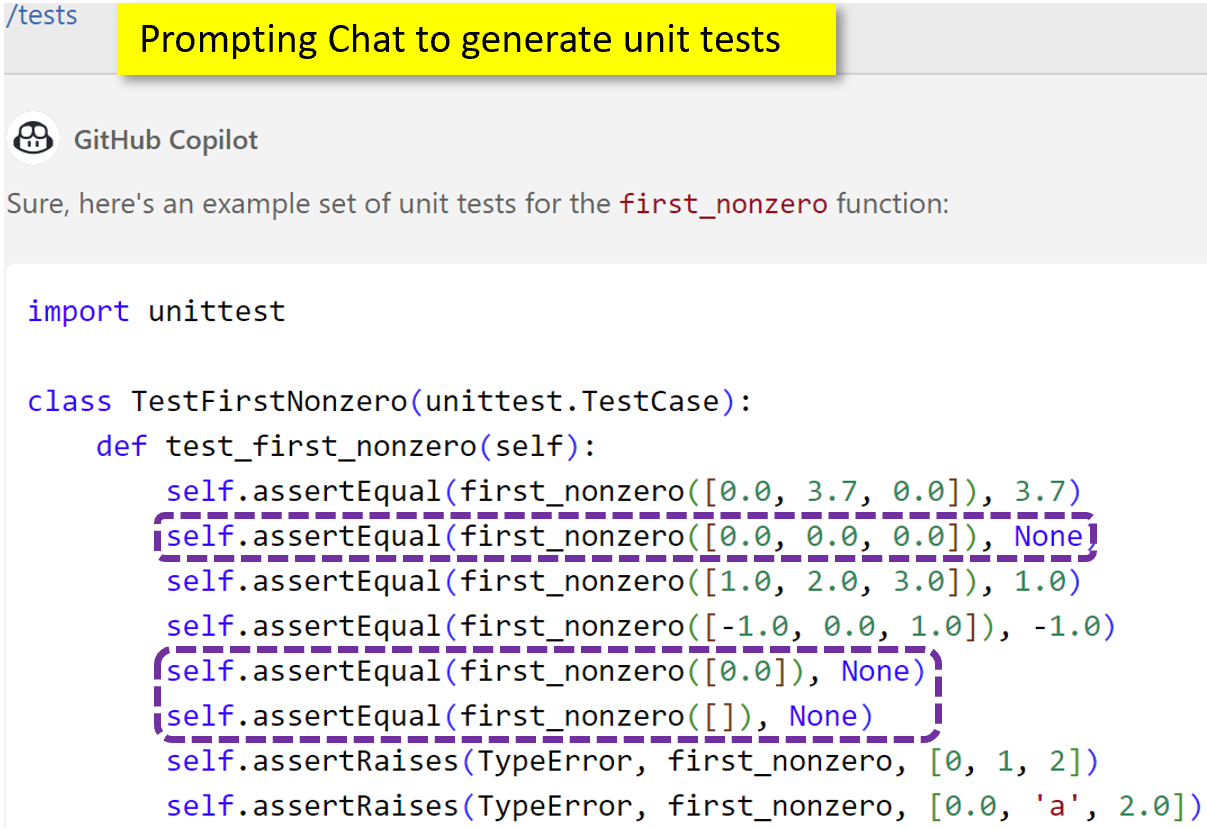}
  \caption{When GitHub Copilot Chat is prompted to generate unit tests, it suggests examples from only one of the two Ambiguous Input Classes (AICs) for this function. For each of these examples (highlighted), Copilot Chat \emph{assumes} that the return value is \texttt{None}.}
  \label{fig:chat}
\end{figure}

In the following Python code, the signature (Line~1) includes a meaningful function name and type-hints\footnote{\url{https://docs.python.org/3/library/typing.html}} for its argument and return type. Further, the purpose statement (Line~2) is expressed as a \emph{docstring}, with one functional example as a \emph{doctest}\footnote{\url{https://docs.python.org/3/library/doctest.html}} (Lines 4-5):

\begin{lstlisting}[language=Python,basicstyle=\ttfamily\small,breaklines=true,postbreak=\mbox{$\hookrightarrow$\space}]
def first_nonzero(nums: list[float]) -> float:
    """Return the first non-zero value in nums.
    
    >>> first_nonzero([0.0, 3.7, 0.0])
    3.7
    """
\end{lstlisting}

This purpose statement is ambiguous. For the class of inputs containing no non-zero values (e.g., \verb|nums = []|), it is unclear what the function should do. When we used GitHub Copilot to generate multiple solutions with the above prompt, it resolved this ambiguity in three ways\footnote{LLMs are usually not deterministic~\cite{chen2021} and can generate different solutions for the same prompt when executed multiple times.} (see Figure~\ref{fig:example-usage}). The first implementation raises an error when \verb|nums| contains no non-zero values\footnote{The third implementation raises a different type of error on such inputs.}:

\begin{lstlisting}[language=Python,basicstyle=\ttfamily\small,firstnumber=7,breaklines=true,postbreak=\mbox{$\hookrightarrow$\space}]
    for num in nums:
        if num != 0.0:
            return num
    raise ValueError("No non-zero numbers in the list")
\end{lstlisting}

The second implementation resolves the ambiguity differently, by replacing Line~10 of the above solution with \verb|return 0.0|. Our tool, GuardRails, cannot determine which of these (if any) is the intended behaviour, but it can at least alert the programmer about this ambiguity by suggesting the input \verb|[]|\footnote{Our tool attempts to report the \emph{simplest} such input.}. Such an input can also be suggested by Copilot's Chat feature, by prompting it to generate tests\footnote{\href{https://docs.github.com/en/copilot/github-copilot-chat/about-github-copilot-chat\#generating-unit-test-cases}{\texttt{docs.github.com/en/copilot/github-copilot-chat/about-github-copilot-chat}}} for this function (Figure~\ref{fig:chat}). Notice that these tests are based on an \emph{assumption} that the expected return value is \verb|None|. Copilot Chat does not explicitly draw attention to potential ambiguities by highlighting these assumptions, so it is up to the programmer to recognize the presence of assumptions.

If this was the \emph{only} input suggested to the programmer, they might attempt to clarify the purpose statement by including an assumption: \verb|nums| will contain at least one non-zero value. However, our tool (but not Copilot Chat) identifies an example from a second class of inputs that exposes a subtler ambiguity: \verb|[nan]|. This list contains ``not a number'', an easy-to-overlook non-zero value whose special properties ensure that the two implementations are once again inequivalent on this input (since \verb|nan != nan|~\cite{ieee754}). Using GuardRails' suggestions, the programmer might clarify ambiguities in the original purpose statement as shown below (Lines 2-4). Lines 9-12 show GitHub Copilot's first suggested implementation.

\begin{lstlisting}[language=Python,basicstyle=\ttfamily\small,breaklines=true,postbreak=\mbox{$\hookrightarrow$\space}]
def first_nonzero(nums: list[float]) -> float:
    """Return the first non-zero value,
    excluding NaN, in nums. If no such
    value exists, return 0.0.
    
    >>> first_nonzero([0.0, 3.7, 0.0])
    3.7
    """
    for num in nums:
        if num != 0.0 and not math.isnan(num):
            return num
    return 0.0
\end{lstlisting}

Our implementation of GuardRails is based on a heuristic, which we detail in Section~\ref{sec:implementation}.

\subsection{Research Questions}
To evaluate our heuristic, we have created a dataset of 15 functions, each with between 1 and 3 ambiguous input classes (AICs). The original version of the \texttt{first\_nonzero()} function belongs to this dataset, and it has two AICs: ``Non non-zero numbers'' and ``NaN as the only non-zero number''. Unlike our heuristic, Copilot Chat does not explicitly highlight certain inputs as potentially ambiguous. Nevertheless, since GitHub Copilot uses a state-of-the-art LLM and is increasingly being adopted by the professional software development community\footnote{\href{https://github.blog/2023-06-27-the-economic-impact-of-the-ai-powered-developer-lifecycle-and-lessons-from-github-copilot/}{\texttt{github.blog/2023-06-27-the-economic-impact-of-the-ai-powered-developer-lifecycle-and-lessons- from-github-copilot}}}, we believe that it provides a good benchmark for comparison with our tool. We first compare the abilities of Copilot Chat and GuardRails across multiple problems:

\begin{description}
\item[RQ1] In relative terms, how do the abilities of both tools to suggest inputs from known AICs vary across the functions in our dataset?
\end{description}

Further, since LLMs are sensitive to the level of detail provided, we expect both these LLM-based techniques to leverage additional details to identify ambiguities in purpose statement. We consider four variants of each function, with progressively increasing amounts of detail:

\begin{description}
\item[S] Only the function's \emph{Signature}.
\item[SP] In addition to \textbf{S}, the (ambiguous) \emph{Purpose statement}.
\item[SP1] In addition to \textbf{SP}, \emph{one} functional example that does not correspond to any AIC.
\item[SPx] In addition to \textbf{SP1}, one or more functional examples that further \emph{explore} the input space, but again do not correspond to any AIC.
\end{description}

The original version of the \verb|first_nonzero()| function presented in this paper is the \textbf{SP1} variant with one functional example for the list \verb|[0.0, 3.7, 0.0]|. The \textbf{SPx} variant adds a functional example for the list \verb|[-3.14]|, which explores parts of the input space that include negative numbers and are ``closer'' to the empty list. We included \textbf{SPx} variants in our study after realising that LLMs can be prompted to generate a richer variety of solutions when such inputs are included. Our second research question is:

\begin{description}
\item[RQ2] In absolute terms, does the percentage of inputs from known AICs increase as we move from \textbf{S} to \textbf{SPx} variants?
\end{description}

\section{Related Work}
Real-world problems are often poorly specified, and the failure of programmers to clarify crucial details before writing code is a known root cause of software project failure~\cite{implicit2022}. Although instructors in CS1 courses often assist novice programmers by providing unrealistically detailed problem specifications, prior work has established that some students fail to comprehend these details and are less likely to produce correct code~\cite{prather2019first, silva2022}. In contrast, we echo Schneider's view~\cite{schneider1978} that CS1 students need greater exposure to problem specifications that are realistic in the sense of containing ambiguities or lacking key details. We believe that such exposure is particularly relevant at a time when 
LLMs such as Codex~\cite{chen2021} can outperform a majority of students \cite{finnie2022robots} on CS1 tasks~\cite{wermelinger2023using,denny2022conversing}.

While several approaches have been proposed to help students comprehend well-specified problems~\cite{lin2021pdl, prather2019first, icer2019, wrenn2020koli}, including approaches suitable for linguistically diverse countries such as India~\cite{t4e2015}, we are unaware of prior work that explicitly points out ambiguities in a given problem. As noted previously, Copilot Chat can implicitly indicate the presence of ambiguities by generating unit tests based on \emph{unacknowledged} assumptions. An aspect of our approach is similar to the Test-Driven User-Intent Formalization workflow proposed by Lahiri et al.~\cite{lahiri2022TDUIF}. Here, the programmer seeks to clarify their intent by providing functional examples, and the LLM is \emph{constrained} to generate code that satisfies these examples. GuardRails similarly uses functional examples to filter out certain implementations (see Section~\ref{sec:implementation}).

LLMs have been used to support CS education in a variety of ways, including explaining code~\cite{macneil2022experiences}, creating code-writing tasks~\cite{macneil2022experiences, sarsa2022automatic}, providing students with hints on such tasks~\cite{pankiewicz2023large}, including how to fix syntactic and logical errors~\cite{leinonen2022using, balse2023iticse}. It is important to note that the output of LLMs is not always correct. For example, LLMs can generate incorrect code~\cite{denny2022conversing} and incorrect explanations~\cite{sarsa2022automatic, balse2023iticse}.

To the best of our knowledge, the approach used by Copilot Chat to generate unit tests uses \emph{only} an appropriately trained LLM. In contrast, our approach uses LLMs to generate multiple implementations and analyses these using two ideas from software testing: property-based testing~\cite{fink1997property} and mutation testing (see~\cite{papadakis2019mutation} for a recent survey). For property-based testing, our implementation uses Python's Hypothesis~\cite{maciver2019hypothesis} library.

\section{Heuristic and Implementation}
\label{sec:implementation}
Our heuristic is based on two key ideas:
\begin{itemize}
    \item If an LLM is given an ambiguous purpose statement for a function and then prompted to generate multiple implementations, it \emph{may} generate two or more \emph{functionally inequivalent} implementations.
    \item Inputs which demonstrate that two implementations are functionally inequivalent \emph{may} reveal ambiguities in the purpose statement.
\end{itemize}

\begin{figure*}
    \includegraphics[width=\linewidth]{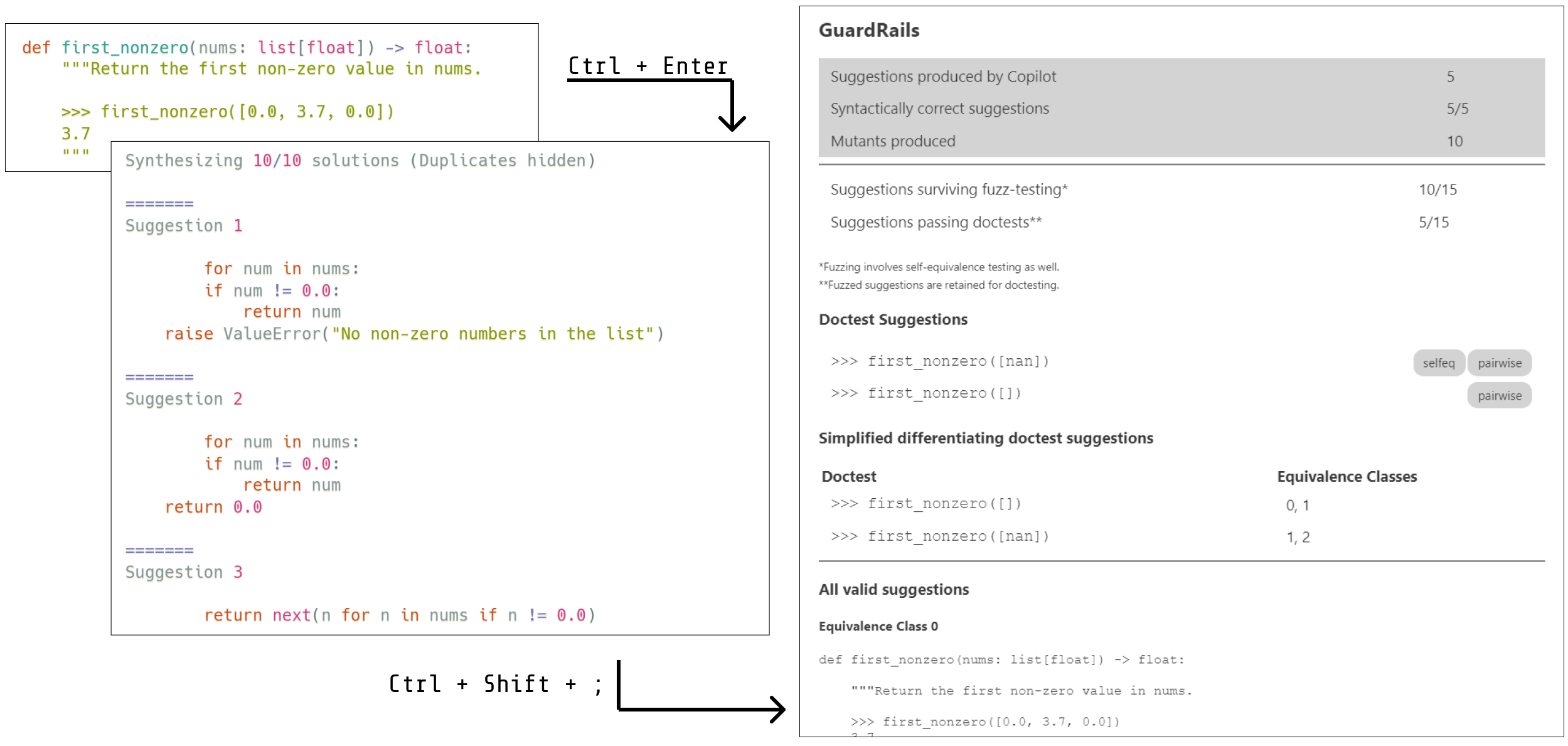}
    \caption{An illustration of our heuristic and implementation for the \texttt{first\_nonzero()} function.}
    \label{fig:example-usage}
\end{figure*}

Our use of the word \emph{may} reflects our uncertainty that the heuristic we have developed on the basis of these ideas will be effective in practice. We defer this concern to Section~\ref{sec:results}. For now, we describe our heuristic and its implementation in GuardRails in detail. The input to our heuristic is a function's signature which specifies the type of each argument, an optional purpose statement that may be ambiguous, and zero or more functional examples. We illustrate our heuristic for the example presented in Section~\ref{sec:motivation}, as shown in Figure~\ref{fig:example-usage}.

\begin{enumerate}
\item Based on the inputs given, we use an LLM to suggest an initial set of syntactically valid implementations for the function. GuardRails works on suggestions provided by GitHub Copilot, which is available as an extension in Visual Studio Code, a popular IDE. Besides inline suggestions, Copilot also provides an option to view the top 10\footnote{This number can be changed in the configurations.} suggestions in a panel. GuardRails picks up the suggestions from this panel and then uses them in its functionality. When it is invoked, GuardRails retains only syntactically valid implementations in a \textit{suggestion space}. In Figure~\ref{fig:example-usage}, this suggestion space initially contains five implementations, three of which are shown.

\item We augment this suggestion space by mutating each implementation. In GuardRails, we use a modified fork of the mutation testing tool MutPy\footnote{\url{https://github.com/mrigankpawagi/mutpy}} to mutate all the initial implementations. In Figure~\ref{fig:example-usage}, we add 10~mutants, resulting in a suggestion space of 15~implementations.

\item We attempt to fuzz~\cite{takanen2018fuzzing} each implementation in the suggestion space by executing it on multiple inputs, as per the function's signature. Fuzzing can cause some implementations to fail on certain inputs (e.g., a division or modulus operation may fail when the second operand is \texttt{0}, or a list \verb|max()| operation may fail on an empty list). In Figure~\ref{fig:example-usage}, we discover the input \verb|[]| during this fuzzing step. In GuardRails, we use Hypothesis' Ghostwriter module to automatically generate a strategy for passing inputs to fuzz each implementation\footnote{\url{https://hypothesis.readthedocs.io/en/latest/ghostwriter.html}}. To ensure that Hypothesis generates inputs of the appropriate type, we require the function's signature to contain type hints (also known as type annotations). We record any inputs that cause such an implement to fail. Some implementations fail because they exceed GuardRails's upper-limit on the execution time (currently, 0.3~seconds per test). This occurs because syntactically valid suggestions may contain infinite loops or may implement an algorithm that is inefficient for some inputs generated by Hypothesis.

\item If some functional examples have been provided, we discard all LLM-generated implementations that fail on any of these examples. For GuardRails, we assume that functional examples have been specified as doctests. We use Python's \verb|doctest| module to discard implementations that fail one or more doctests, thereby trimming the suggestion space. In Figure~\ref{fig:example-usage}, this trims the suggestion space from 15~implementations to just 5~implementations.

\item For each pair of implementations in the resulting suggestion space, we attempt to find an input on which these implementations are functionally inequivalent. In GuardRails, we create pair-wise equivalence tests and use the input strategy created in Step~(3) to run these tests using Hypothesis, for finding such an input. Hypothesis uses heuristics to \emph{shrink} failing inputs\footnote{\url{https://hypothesis.readthedocs.io/en/latest/data.html\#shrinking}}, often resulting in simple, easy-to-read tests. In Figure~\ref{fig:example-usage}, we discover the input \verb|[nan]| during this step.

\item Finally, we collate all recorded inputs and present these to the programmer. In GuardRails, we report these as \emph{partial} doctests i.e., we specify the input and explicitly prompt the programmer to provide the expected output (Figure~\ref{fig:example-usage}). This is in contrast to Copilot Chat, which assumes an expected output without revealing the potential ambiguity (Figure~\ref{fig:chat}).
\end{enumerate}

GuardRails is open-sourced and is available as a GitHub repository\footnote{\url{https://github.com/mrigankpawagi/GuardRails}}. It is implemented as a VSCode Extension, a link to which can be found in the repository along with a link to the performance comparison dataset discussed in Section~\ref{sec:results}. Once installed in VSCode, GuardRails can be invoked by first triggering Copilot's suggestions panel, and then pressing an appropriate key combination\footnote{\texttt{Ctrl+Shift+;} runs the full heuristic as described here. \texttt{Ctrl+Shift+/} skips step~(2) i.e., it does not generate any mutants. For some problems, MutPy generates an excessive number of mutants, resulting in degraded performance.}. 

\section{Comparison with Copilot Chat}
\label{sec:results}
To the best of our knowledge, Copilot's Chat feature relies \emph{only} on a suitably trained LLM when prompted to generate unit tests (Figure~\ref{fig:chat}). In contrast, GuardRails prompts the underlying LLM to generate code, and this code is evaluated using additional tools (the doctest and Hypothesis modules, and MutPy). We expect this additional computation to result in an improved ability to identify inputs from AICs. Both tools rely on LLMs, whose results are not deterministic~\cite{chen2021}. Hence, for each (function, variant) combination, we executed both tools 5~times and we report the best result (top@5).

\subsection{RQ1: Relative Performance}
\begin{figure}
\begin{minipage}[t]{0.48\textwidth}

  \includegraphics[width=1\linewidth]{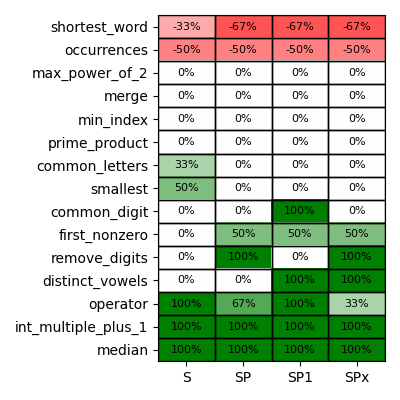}
    \caption{Differences in percentages of AIC (for each variant of all 15-questions) caught by GuardRails and GitHub Copilot Chat (top@5).}
    \label{fig:gr-cc-heatmap}
\end{minipage}\hfill
\begin{minipage}[t]{0.48\textwidth}
  \includegraphics[width=1\linewidth]{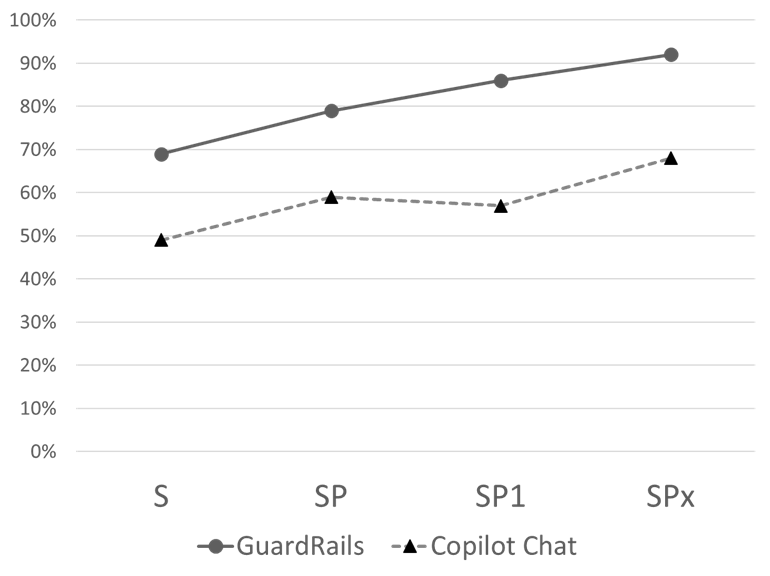}
  \caption{The percentage of AICs (averaged over all 15 questions) found by GitHub Copilot Chat vs. GuardRails (top@5).}
  \label{fig:gr-cc}
\end{minipage}
\end{figure}

A comparison of the abilities of Copilot Chat relative to GuardRails is shown in Figure~\ref{fig:gr-cc-heatmap}, for the 15~functions and their 4~variants. The numbers within each row are similar, indicating that differences in relative performance are due to the specifics of each function. Figure~\ref{fig:gr-cc-heatmap} shows that Copilot Chat identifies inputs from more AICs than GuardRails for 2/15 functions (red), the performance of the two tools is similar for 7/15 functions (largely white), and GuardRails is better for 6/15 functions (largely green).

\subsection{RQ2: Absolute Performance by Variant}
When we average the performance across all functions, we find that both tools are able to leverage increasing levels of detail to a similar extent (Figure~\ref{fig:gr-cc}). However, while Copilot Chat is able to raise the average percentage of AICs found from 49\% (variant \textbf{S}) to 68\% (variant \textbf{SPx}), GuardRails starts from a higher base of 69\% (variant \textbf{S}) and improves to 93\% (variant \textbf{SPx}).

\section{Limitations}
\label{sec:limitations}
Although the results of our heuristic are promising, we acknowledge that GuardRails is a research prototype with several limitations. We believe that these limitations can be addressed so that our heuristic can be incorporated into professionally developed tools such as Copilot Chat. Our key limitations are:

\begin{enumerate}
\item We have focused on making GuardRails usable for only one programming language (Python~3) in the specific context of \emph{individual} functions. This prevents our tool from being used for whole programs, or even for functions that call helper functions (e.g., functions imported from other modules). Further, to ensure that Hypothesis can ghostwrite accurate tests, functions must have type hinting.

\item Since GuardRails relies on the underlying LLM to generate complete implementations, it is presently suitable for simple problems (e.g., CS1 level problems). As LLMs continue to improve, our heuristic could be applicable for more complex problems~\cite{li2022competition, openai2023gpt4}.

\item Since key components of GuardRails are not deterministic (LLMs~\cite{chen2021} and Hypothesis~\cite{maciver2019hypothesis}) our tool occasionally produces poor results. A similar limitation applies to Copilot Chat, which is why we report top@5 results in Section~\ref{sec:results}.
\end{enumerate}

\section{Discussion and Future Work}
\subsection{Usage by Instructors}
We believe that instructors will find GuardRails useful in two ways. First, while creating code-writing tasks (e.g., for high-stakes examinations), instructors can use our tool to check problem statements for ambiguities. We believe that it would be particularly interesting to investigate the utility of GuardRails for this purpose in a linguistically diverse country such as India. Second, instructors may wish to deliberately write ambiguous problem statements~\cite{schneider1978}. In this case, GuardRails can be used to confirm that the ambiguities in the specification are the ones expected.

\subsection{Usage by Novice Programmers}
We have demonstrated principled ways in which even novice programmers can increase the level of detail presented to LLMs (from variant \textbf{S} to variant \textbf{SPx}) in order to improve their ability to detect ambiguities. As we have acknowledged in Section~\ref{sec:limitations}, GuardRails can be used in limited contexts. We hypothesize that exposure to ambiguities highlighted by our tool can help novices develop the ability to identify ambiguities in broader contexts as well. This line of research seems particularly promising for students with limited fluency in the language in which problems are specified.

\section{Conclusions}
We have proposed a novel heuristic that uses LLMs to identify potential ambiguities in the purpose statements of Python functions. Further, we compared our open-source tool, GuardRails, against a production-level benchmark (Copilot Chat). We have demonstrated that our tool can \emph{explicitly} identify potential ambiguities and is often (but not always) able to outperform Copilot Chat. We hope that the ideas presented here are incorporated into widely used, professionally developed tools such as GitHub Copilot. We believe our heuristic can further enhance the productivity of software developers and also empower novice programmers. Finally, we believe that our heuristic can support new approaches to CS pedagogy and assessment that expose students to deliberately ambiguous problem specifications.

\bibliographystyle{unsrt}  
\bibliography{arxiv}

\end{document}